\documentclass[11pt]{article}
\usepackage{geometry} 
\usepackage[parfill]{parskip}    % Activate to begin paragraphs with an empty line
\usepackage{graphicx}
\usepackage{amssymb}
\usepackage{epstopdf}
      \makeatletter
      
      \@addtoreset{equation}{section}
      \makeatother
\newcommand{\emp}[1]{{\bf #1}}

\newtheorem{teiri}{\emp{Theorem}}[section]
\newtheorem{kei}{\emp{Corollary}}[section]

\newtheorem{hodai}{\emp{Lemma}}[section]
\newtheorem{teigi}{\emp{Definition}}[section]
\newtheorem{rei}{\emp{Example}}[section]
\newtheorem{chui}{\emp{Remark}}[section]

\newcommand{\QED}{\hspace*{\fill}$\Box$ }

\newcommand{\spr}[1]{{\bf #1}}

\newcommand{\vep}{\varepsilon}
\newcommand{\vph}{\varphi}

\newcommand{\cC}{{\cal C}}
\newcommand{\cD}{{\cal D}}

\newcommand{\cM}{{\cal M}}
\newcommand{\cN}{{\cal N}} %ch7

\newcommand{\cR}{{\cal R}} %ch7

\newcommand{\cX}{{\cal X}}
\newcommand{\cY}{{\cal Y}}

\newcommand{\ssx}{\spr{x}}

\newcommand{\nth}{\frac{1}{n}}

\newcommand{\bteiri}{\begin{teiri}}
\newcommand{\eteiri}{\end{teiri}}
\newcommand{\bkei}{\begin{kei}}
\newcommand{\ekei}{\end{kei}}
\newcommand{\brei}{\begin{rei}}
\newcommand{\erei}{\end{rei}}
\newcommand{\bhodai}{\begin{hodai}}
\newcommand{\ehodai}{\end{hodai}}
\newcommand{\bteigi}{\begin{teigi}}
\newcommand{\eteigi}{\end{teigi}}
\newcommand{\bchui}{\begin{chui}}
\newcommand{\echui}{\end{chui}}
\newcommand{\beq}{\begin{equation}}
\newcommand{\eeq}{\end{equation}}
\newcommand{\beqn}{\begin{eqnarray}}
\newcommand{\eeqn}{\end{eqnarray}}
\newcommand{\beqns}{\begin{eqnarray*}}
\newcommand{\eeqns}{\end{eqnarray*}}

\newcommand{\map}{\vph_n: \cX^n \to \cY^n}
\newcommand{\mapMtoY}{\vph_n: \cM_{M_n} \to \cY^n}
\newcommand{\mapXtoM}{\vph_n: \cX^n \to \cM_{M_n}}

\title{~\\~\\~\\~\\ \bf Multicasting  Correlated Multiple  Sources to Multiple  Sinks over a Noisy Network\thanks
{Presented at International Symposium on Information Theory and its Applications,
 Auckland, New Zealand, Dec. 7-10, 2008}
}
\author{~\\~\\~\\~\\Te Sun HAN\thanks{Te Sun Han is with 
Faculty of Science and Engineering, Waseda University, 
Okubo 2-4-12-902, Shinjuku-ku
Tokyo 169-0072, Japan. 
E-mail:\ han@is.uec.ac.jp,\ han@aoni.waseda.jp}}

\date{\today}
\begin{document}
\setcounter{page}{0}
\maketitle
\thispagestyle{empty}
\newpage

\pagenumbering{roman}
%\pagestyle{roman}

%
%Dec.26, 1996: Pictures were inserted.
%
%\pagestyle{headings}

\pagenumbering{arabic}
\setcounter{page}{0}
\setcounter{equation}{0}

\vspace*{3cm}

{\bf Abstract:} \ The problem of network coding for multicasting  a single source to multiple  sinks has first been studied by Ahlswede, Cai, Li and Yeung in 2000, in which they have established the celebrated max-flow mini-cut theorem on non-physical information flow over a network of independent channels. 
On the other hand, in 1980, Han has studied the case with correlated multiple  sources and a single sink from the viewpoint of polymatroidal functions in which a necessary and sufficient condition has been demonstrated for reliable transmission over the network. This  paper presents an attempt
to unify both cases, which leads to establish a necessary and sufficient condition for reliable 
transmission over a network for multicasting  correlated  multiple  sources to multiple  sinks. Here, the problem of separation of source coding and network coding is also discussed.

\vspace{1cm}         

{\bf Index terms:} \ network coding, multiple  sources, multiple  sinks, correlated sources, entropy rate,
capacity function, polymatroid, co-polymatroid, mini-cut, transmissibility
\newpage

%
%
%\large
%
%section4.0
\section{Introduction}\label{intro-geri1}
The problem of network coding for multicasting  a single source to multiple  sinks has first been studied by Ahlswede, Cai, Li and Yeung \cite{al-yeung}  in 2000, in which they have established the celebrated max-flow mini-cut theorem on non-physical information flow over a network of independent channels. 
On the other hand, in 1980, Han \cite{han-cover} had studied the case with {\em correlated} multiple  sources and a single sink from the viewpoint of polymatroidal functions in which a necessary and sufficient condition has been demonstrated for reliable transmission over a  network. 

This  paper presents an attempt
to unify both cases and to generalize it to  quite a general case with stationary ergodic correlated sources and {\em noisy} channels (with {\em arbitrary} nonnegative real values of capacity that are {\em not} necessarily {\em integers}) satisfying the strong converse property (cf. Verd\'u and Han \cite{verdu-han}, Han \cite{han-book}),
which leads to establish a necessary and sufficient condition for reliable 
transmission over a  {\em noisy} network for multicasting  correlated  multiple  sources altogether to every multiple  sinks. 

It should be noted here that 
in such a   situation with {\em correlated} multiple  sources, the central issue turns out to be how to construct the 
{\em matching condition} between source and channel (i.e., joint source-channel coding), instead of  of  the traditional concept of {\em 
capacity region} (i.e., channel coding), although in the special case with {\em non-correlated} independent multiple sources the problem reduces again to how to describe the capacity region.
%domain

Several network models with correlated multiple  sources have been studied by some people, e.g.,
by Barros and Servetto \cite{bar-serv}, Ho, M\'edard, Effros and Koetter \cite{koetter}, 
Ho, M\'edard,  Koetter, Karger, Effros, Shi and Leong \cite{koetter-1}, 
Ramamoorthy, Jain, Chou and Effros \cite{effros}.
Among others, \cite{koetter}, \cite{koetter-1} and \cite{effros} consider (without attention to the {\em converse} part) a very  restrictive case of {\em error-free} network coding 
for two  stationary memoryless  correlated sources with a single sink to study
the error exponent problem,
where we notice that 
 %it is assumed  in \cite{koetter} and  \cite{koetter-1} 
 %(\cite{koetter} is included in \cite{koetter-1}) that 
% the paths from source nodes to sink nodes are  all of the same length $L$;  and 
 %
  all the arguments in \cite{koetter}, \cite{koetter-1} and \cite{effros}
 % (\cite{koetter} is included in \cite{koetter-1})
  can be validated  only  within  the narrow class of stationary memoryless sources of {\em  integer} bit rates and 
 {\em error-free} channels (i.e., the {\em identity} mappings) all with {\em one bit} (or {\em integer bits}) capacity
   (these restrictions are needed solely to invoke ``Menger's theorem" in graph theory). 
   The main result in the present paper is quite free from such severe restrictions, because we can dispense with the use of Menger's theorem.

On the other hand,  \cite{bar-serv} revisits the same model as in Han \cite{han-cover}, while
 \cite{effros} focuses on  the network with two correlated sources and two sinks to discuss 
the separation problem of distributed source coding (based on Slepian-Wolf theorem) and network coding. It should be noted that, in the case of networks with correlated multiple  sources, such a {\em separation} problem is another central issue,  although it is yet far from fully solved.
%in addition to the matching condition problem. 
%(also, cf. Song, Yeung and Cai \cite{song}). 
In this paper, we mention  a  sufficient
condition for separability in the case with multiple sources and multiple sinks.
%of distributed source coding and network coding 
(cf. Remark \ref{chui:geriko1}). 
% can be written in terms of combinatorial polymatroids.

On the other hand, we may consider another network model with {\em independent} multiple  sources
but with multiple  sinks each of which is required to reliably reproduce a prescribed subset of the 
multiple  sources that depends on each sink. However, the problem with this general model looks quite hard, although, e.g., 
 Yan, Yeung and Zhang \cite{yan} and  Song, Yeung and Cai \cite{song-yeung} 
have demonstrated the entropy  characterizations of the capacity region, which  still contain 
 limiting operations and are not computable.
Incidentally, Yan, Yang and Zhang \cite{zhang} have considered, as a computable special case,  degree-2 three-layer networks with 
$K$-pairs transmission requirements to derive the explicit capacity region.
 In this paper, for the same reason, we  focus on the case in which all the correlated multiple sources 
is to be multicast to all the multiple sinks  and  derive a simple necessary and sufficient matching condition
in terms of conditional entropy rates and capacity functions. This case can be regarded as   the network counterpart of the non-network compound Slepian-Wolf system \cite{csis-kor}.

 We notice here the following; although throughout in the paper we are encountered with the subtleties coming from the general channel and source characteristics assumed, the main logical stream remains essentially  unchanged
 if  we consider  simpler models, e.g.,  such as stationary  correlated Markov sources
 together with stationary memoryless noisy channels. This means that considering only simple cases
 does not help so much at both of the conceptual and  notational levels of the arguments.
 For this reason, we preferred here the compact general settings.
 
The present paper consists of five sections. In Section \ref{ss:HT_LD} notations and preliminaries are described, and 
in Section \ref{ss:HT_LD_C} we state the main result as well as its proof.
In Section \ref{ss:geri-ex1} two examples are shown. Section \ref{alter-cond} provides another type of necessary and sufficient condition for transmissibility.
Finally, some detailed comments on the previous papers are given.
%
%section4.4
\section{Preliminaries and Notations \label{ss:HT_LD}}

\medskip

\noindent
\qquad{\em A. Communication networks} 

\medskip

Let us consider an acyclic directed graph $G=(V,E)$ where 
$V =\{1,2,\cdots, |V|\}$ $
(|V| <+\infty)$, $E\subset V \times V,$ but
$(i,i)\not\in E$ for all $i\in V$. Here, elements of $V$ are called {\em nodes}, and
elements $(i,j)$ of $E$  are called {\em edges} or {\em channels}
from $i$ to $j$. Each edge $ (i,j)$ is assigned the {\em capacity}  $c_{ij}\ge 0$,
which specifies the maximum amount of information flow passing through the
channel $(i, j)$. If we want to emphasize the graph thus capacitated, we write it as $G=(V, E,C)$ where $C=(c_{ij})_{(i,j)\in E}$. A graph $G=(V, E,C)$ is sometimes called a (communication) network, and indicated also by $\cN=(V, E,C)$.
We consider two fixed subsets $\Phi, \Psi$ of $V$ such that 
$\Phi \cap \Psi = \emptyset$  (the empty set) with
\[
\Phi = \{s_1, s_2, \cdots, s_p\},
\]
\[
\Psi = \{t_1, t_2, \cdots, t_q\},
\]
where  elements of $\Phi$ are called  {\em source nodes}, while elements of $\Psi$ are called {\em sink nodes}.
Here, to avoid subtle irregularities, we assume that there are no edges $(i,s)$ such that $s \in \Phi.$

Informally, our problem is  how to simultaneously transmit the information generated at the source nodes in $\Phi$ altogether to all the sink nodes in $\Psi$.
More formally, this problem is described as in the following subsection.
\bchui\label{chui:tumari1}
{\rm
In the above we have assumed that $\Phi\cap \Psi = \emptyset$. However, we can reduce the case of 
$\Phi\cap \Psi \neq \emptyset$ to the case of $\Phi\cap \Psi = \emptyset$ by equivalently modifying
the given network. In fact, suppose $\Phi\cap \Psi \neq \emptyset$ and let $k\in 
\Phi\cap \Psi$ for some $k$. Then, we add a new source node $k^{\prime}$ to
$\Phi$, and generate a new edge $(k^{\prime}, k)$  with capacity $\infty$, and remove the node $k$ from $\Phi$. Repeat this procedure until we have $\Phi\cap \Psi = \emptyset$. The assumption 
that there are no edges $(i,s)$ such that $s \in \Phi$ also can be dispensed with by repeating a
similar procedure.
} \QED
\echui

\medskip

\noindent
\qquad{\em B. Sources and channels} 

\medskip

Each source node $s \in \Phi$  generates a stationary and  ergodic source process
\beq\label{eq:geri2}
X_s = (X_s^{(1)}, X_s^{(2)}, \cdots),
\eeq
where $X_s^{(i)} (i=1,2,\cdots)$  takes values in finite source alphabet $\cX_s$.
Throughout in this paper we consider the case in which the whole joint process $X_{\Phi} \equiv (X_s)_{s \in \Phi}$ 
is  stationary and ergodic.
It is then evident that the joint process $X_{T} \equiv (X_s)_{s \in T}$ is also stationary and ergodic for 
any $T$ such that $\emptyset \neq T \subset \Phi $.
The component processes $X_s\  (s\in \Phi) $  may be correlated.
We write $X_T$ as 
\beq\label{eq:geri3}
X_T = (X_T^{(1)}, X_T^{(2)}, \cdots)
\eeq
and put
\beq\label{eq:geri4}
X_T^n = (X_T^{(1)}, X_T^{(2)}, \cdots, X_T^{(n)}),
\eeq
where $X_T^{(i)}\  (i=1,2,\cdots)$ takes values in $ \cX_T\equiv \prod_{s\in T}\cX_s$.

% let us assgn a finite source alphabet $\cX_{s}$
%and generate a source (random) variable $X_{s}$
%taking values in $\cX_{s}$ where $X_{s}$'s $\in \Phi$ may be correlated,
%subject to joint probability $p_{\Phi}(x_{s_1},\cdots, x_{s_p})$.
%We denote by $X^n_{\Phi}$ an $n$ independent copies of 
%$X_{\Phi}=(X_{s_1},\cdots, X_{s_p})$. (More generally, 
%they may be stationary and ergodic sources %of block length $n$.)
% Incidentally, we write an $n$ independent copies 
%of $X_S =(X_m)_{m \in S}$ as $X^n_{ S}$\ ($\emptyset \neq S \subset \Phi)$.

%\smallskip

On the other hand, it is assumed that all the channels $(i,j)\in E,$ specified by the transition probabilities 
%
%Accordingly, as we are interested in the asymptotics of network coding,
% it may be assumed that the channel $(i, j)\in E$, specified by the transition probabilities
$w_{ij}: A_{ij}^n \to B_{ij}^n$ with finite input alphabet $A_{ij}$ and
finite output alphabet $B_{ij}$, 
 are {\em statistically  independent} and satisfy the {\em strong converse } property (see Verd\'u and Han \cite{verdu-han}). It should be noted here that stationaty and memoryless ({\em noisy} or {\em noiseless}) channels with finite
 input/output alphabets satisfy, as  very special cases,  this property (cf. Gallager \cite{gall}, Han \cite{han-book}).
 Barros and Servetto \cite{bar-serv} have considered the case of stationary and  memoryless sources/channels with finite alphabets.
 The following lemma plays a crucial role in establishing the relevant converse of the main result:
 
 \bhodai\label{hoda-strong-cap}{\rm (Verd\'u and Han \cite{verdu-han})}
{\rm 
The channel capacity $c_{ij}$ of a channel $w_{ij}$ satisfying the strong converse property with finite input/output alphabets is given by
\[
c_{ij} = \lim_{n\to\infty} \nth \max_{X^n}I(X^n;Y^n),
\]
where $X^n, Y^n$ are the input and the output of the channel $w_{ij}$, respectively, and 
$I(X^n; Y^n)$ is the mutual information (cf. Cover and Thomas \cite{cover-thomas}).
\QED}
\ehodai

\medskip

\noindent
\qquad{\em C. Encoding and decoding} 

\medskip

In this section let us state the necessary operation of encoding and decoding for network coding with correlated multiple sources to be multicast to multiple sinks. 

With  arbitrarily
small $\delta>0$ and   $\vep>0$, we introduce an 
$(n, (R_{ij})_{(i,j)\in E},$ $
\delta, \vep )
$ code as the one as specified by (\ref{newji-1}) $\sim$ (\ref{newji-7}) below, where we use the notation $[1, M]$ to indicate $
\{1,2,\cdots, M\}$. How to construct a ``good" $(n, (R_{ij})_{(i,j)\in E},$ $
\delta, \vep )$ code will be shown in Direct part of the proof of Theorem \ref{teiri:newji-1}.

\medskip

\noindent
\quad{\em 1)} For all $(s,  j)\ (s \in \Phi)$, the encoding function is
\beq\label{newji-1}
f_{sj}: \cX^n_s\to [1, 2^{n(R_{sj}-\delta)}],
\eeq
where the output of $f_{sj}$ is carried over to the encoder $\varphi_{sj}$ of
channel $w_{sj}$, while the decoder $\psi_{sj}$ of $w_{sj}$ outputs an estimate of the output of $f_{sj}$, which is specified by the stochastic composite function: 
\beq\label{newji-2}
h_{sj}\equiv \psi_{sj}\circ w_{sj}\circ \varphi_{sj}\circ f_{sj}: \cX^n_s\to [1, 2^{n(R_{sj}-\delta)}];
\eeq
\medskip
\noindent
\quad{\em 2)} For all $(i,  j)$ $ ( i \not\in \Phi)$,
the encoding function is
\beq\label{newji-3}
f_{ij}: \prod_{k:(k,i)\in E} [1, 2^{n(R_{ki}-\delta)}]
\to [1, 2^{n(R_{ij}-\delta)}],
\eeq
where the output of $f_{ij}$ is carried over to the encoder $\varphi_{ij}$ of
channel $w_{ij}$, while the decoder $\psi_{ij}$ of $w_{ij}$ outputs an estimate of the output of $f_{ij}$, which is specified by the stochastic composite function: 
%\beq\label{newji-4}
%h_{sj}\equiv W^n_{sj}\circ f_{sj}: \cX^n_s\to [1, 2^{n(R_{sj}-\delta)}];
%\eeq
\beq\label{newji-4}
h_{ij}\equiv \psi_{ij}\circ w_{ij}\circ \varphi_{ij}\circ f_{ij}: \prod_{k:(k,i)\in E} [1, 2^{n(R_{ki}-\delta)}]
\to [1, 2^{n(R_{ij}-\delta)}].
\eeq
Here, if $\{k:(k,i)\in E\}$ is empty, we use the convention that $f_{ij}$ is an arbitrary constant function taking a value in  $[1, 2^{n(R_{ij}-\delta)}]$;
\medskip

\noindent
\quad{\em 3)} For all $t\in \Psi$, the decoding function is 
\beq\label{newji-5}
g_t: \prod_{k:(k,t)\in E} [1, 2^{n(R_{kt}-\delta)}]
\to \cX^n_{\Phi}.
\eeq
\noindent
\quad{\em 4) Error probability} 

\medskip
\noindent
All sink nodes $t\in \Psi$ are required to reproduce  a ``good" estimate  $\hat{X}^n_{\Phi,t}$
($\equiv$ the output of the decoder $g_t$) 
 of $X^n_{\Phi}$, through the network
 $\cN = (V,E,C)$, so that the error probability 
 $\Pr\{\hat{X}^n_{\Phi,t} \neq X^n_{\Phi}\}$ be as small as possible.
Formally, 
for all $t\in \Psi$,  the probability  $\lambda_{n,t}$ of decoding error committed at sink $t$
is required to satisfy
\beq\label{newji-7}
\lambda_{n,t} \equiv \Pr\{\hat{X}^n_{\Phi,t} \neq X^n_{\Phi}\} \le \vep
\eeq
for all sufficiently large $n$.
%where $\hat{X}^n_{\Phi,t}$ is the  output of the decoder $g_t$.
Clearly, $\hat{X}^n_{\Phi,t}$  are the random variables induced by 
$X^n_{\Phi}$ that were generated at all source nodes $s \in \Phi$.

\bchui\label{oldji-1}
{\rm
%In the above, $f_{ij}$ is the encoding function for  edge $(i,j)$, while $g_t$
%is the decoding function for sink $t$.
In the above coding process, $f_{ij}$ is applied before $f_{i^{\prime}{j\prime}}$ is
if $i<i^{\prime}$, and $f_{ij}$ is applied before $f_{i{j\prime}}$ is if
$j < j^{\prime}.$ Such an indexing is possible because we are dealing with {\em acyclic} directed graphs. This defines the order in which the encoding functions are applied. Since  $i < j $ if $(i,j)\in E$, a node does not encode until all the necessary informations are received on the input channels
 (see, Ahlswede, Cai, Li and Yeung \cite{al-yeung}, Yeung \cite{yeng-first}). In this sense, the coding procedure with the codes 
 $(n, (R_{ij})_{(i,j)\in E}, \delta, \vep)$ defined above is in  accordance with the natural ordering on an acyclic graph. This observation will be fully used in the proof of Converse part of Theorem \ref{teiri:newji-1} in order to 
establish a Markov chain property.}
\QED
\echui

We now need the following definitions.

\bteigi [{\it rate achievability}]\label{teigi-newji-1}
{\rm
If there exists an $(n, (R_{ij})_{(i,j)\in E},$ $
\delta, \vep )
$ code  for any arbitrarily small $\vep>0$ as well as any sufficiently small $\delta>0$,
and for all sufficiently large $n$, then we say that the rate 
$(R_{ij})_{(i,j)\in E}$ is {\em achievable} for the network
$G=(V, E)$.
}
\QED
\eteigi
\bteigi [{\it transmissibility}]\label{teigi-newji-2}
{\rm
If, for any small $\tau>0$,  the augmented capacity rate $(R_{ij}=c_{ij}+ \tau)_{(i,j)\in E}$ is
achievable, then we say that the source $X_{\Phi}$ is {\em transmissible} over the network $\cN=(V,E, C),$ where $c_{ij} + \tau $ is called the $\tau$-capacity of channel $(i,j).$
}
\QED 
\eteigi
 The proof  of Theorem \ref{teiri:newji-1} (both of the converse part and the direct part) are based  on these definitions. 
%
%Thus, in the above process of coding, we have
%\beq\label{newj-8}
%h_{ij} = f_{ij} \quad (\forall (i,j)\in E)
%\eeq
%with probability at least $1-|E|\gamma_n \to 1$ ($n \to\infty).$
%}

\medskip

\noindent
\qquad{\em D. $\lambda$-Typical sequences} 
\medskip

Let $\ssx_{\Phi}$ denote the sequence of length $n$ such as
\[
\ssx_{\Phi} = (x_{\Phi}^{(1)}),\cdots, x_{\Phi}^{(n)}) \in \cX_{\Phi}^n.
\]
Similarly, we denote by $\ssx_T \ (\emptyset \neq T \subset \Phi)$ the sequence such as
\[
\ssx_T = (x_T^{(1)}),\cdots, x_T^{(n)}) \in \cX_{T}^n.
\] 
We set 
\[
p(\ssx_T) = \Pr\{ X_T^n = \ssx_T\}
\]
and let $H(X_T)$ be the entropy rate of the process $X_T$. With any small $\lambda >0$,  we say that 
$\ssx_{\Phi} \in \cX^n_{\Phi}$ is a $\lambda$-typical sequence if 
\beq\label{eq:dento1}
\left| \nth \log\frac{1}{p(\ssx_S)} - H(X_{S})\right| < \lambda\quad (\emptyset \neq \forall S \subset \Phi),
\eeq
where $\ssx_{S}$ is the projection of  $\ssx_{\Phi}$  on the $S$-direction, i.e., $\ssx_{\Phi} = (\ssx_{S}, \ssx_{\overline{S}})$  ($\overline{S}$ is the complement of $S$ in $ \Phi$).
 We shall denote by $T_{\lambda}(X_{\Phi})$
the set of all  $\lambda$-typical sequences. For any subset $\emptyset \neq S \subset 
\Phi$, let $T_{\lambda}(X_S)$ denote the projection of  $T_{\lambda}(X_{\Phi})$
on $\cX_S^n$; that is,
\beq\label{newji-15}
T_{\lambda}(X_S) = \{\ssx_S \in \cX_S^n| (\ssx_S, \ssx_{\overline{S}}) \in T_{\lambda}(X_{\Phi})
\ \mbox{for some}\  \ssx_{\overline{S}} \in \cX_{\overline{S}}^n\}.
\eeq
%where $\overline{S}$ is the complement of $S$ in $ \Phi$.
%In general, we use $\ssx_{T}$ to denote sequences taking values in
%$ \cX_T^n $ for a subset $T \subset \Phi $.
%
%
Furthermore, set for any $\ssx_{\overline{S}} \in T_{\lambda}(X_{\overline{S}})$,
\beq\label{newji-16}
T_{\lambda}(X_S | \ssx_{\overline{S}}) = \{\ssx_S \in \cX_S^n|
(\ssx_S, \ssx_{\overline{S}}) \in T_{\lambda}(X_{\Phi})\}.
\eeq
We say that $\ssx_S $ is jointly typical with $\ssx_{\overline{S}}$
if $\ssx_S \in  T_{\lambda}(X_S|\ssx_{\overline{S}})$.
Now we have (e.g., cf. Cover and Thomas \cite{cover-thomas}):
\bhodai\label{nu-hodai} \mbox{}

{\rm
1)\ For any small  $\lambda >0$ and for all sufficiently large $n$,
\beq\label{newji-16-1}
\Pr\{X_{\Phi}^n \in T_{\lambda}(X_{\Phi})\} \ge 1-\lambda;
\eeq
 2) for any $\ssx_{\overline{S}} \in T_{\lambda}(X_{\overline{S}})$,
\beq\label{newji-17}
|T_{\lambda}(X_S | \ssx_{\overline{S}}) | \le 2^{n(H(X_S|X_{\overline{S}}) + 2\lambda)},
\eeq
where $H(X_S|X_{\overline{S}})=H(X_{\Phi}) - H(X_{\overline{S}})$ is the conditional entropy rate (cf. Cover
\cite{cover}). Specifically,
$$
H(X_S|X_{\overline{S}}) = \lim_{n\to\infty} \nth H(X^n_S|X^n_{\overline{S}}).
$$
\QED}
\ehodai 
This lemma will be used in the process of proving the transmissibility of the source 
$X_{\Phi}$ over  the network $\cN = (V,E,C)$.
\medskip

\noindent
\qquad{\em E. Capacity functions}  
\medskip

Let $\cN = (V,E,C)$ be a network. 
For any subset $M\subset V$ we say that $(M, V \setminus M)$ 
(or simply, $M$) is a {\em cut} and 
$$E_M \equiv \{(i,j) \in E | i \in M, j\in V \setminus M\}$$
 the {\em cutset} of $(M, V \setminus M)$
(or simply, of $M$).
Also, we call 
%\beq\label{eq:newji-1}
%c(M,V \setminus M) \equiv \sum_{\begin{array}{c}
%(i,j)\in E \\ i \in M \\ j \in V_M
%\end{array}}
%\eeq
%\beq\label{eq:1}
\beq\label{eq:newji-1}
c(M,V \setminus M) \equiv \sum_{ 
(i,j) \in E ,  i \in M,  j \in V \setminus M} c_{ij}
\eeq
%\eeq
the value of the cut $(M, V \setminus M)$. Moreover, 
for any subset $S$ such that 
$\emptyset \neq S \subset \Phi$ (the source node set) and
for any $t\in \Psi$ (the sink node sets),
define
\beq\label{eq:newji-2}
\rho_t (S)= \min_{M: S \subset  M, t\in V \setminus M} c(M,V \setminus M);
\eeq
\beq\label{eq:newji-34}
\rho_{\cN} (S)= \min_{t\in \Psi} \rho_t (S).
\eeq
We call this $\rho_{\cN} (S)$ the capacity function of
 $S\subset  V$ for the network $\cN = (V,E,C)$.
\bchui\label{chui-aho1}
{\rm
A set function $\sigma(S)$ on $\Phi$ is called a co-polymatroid
\footnote{In Zhang, Chen, Wicker and Berger \cite{xzhang}, 
the {\em co}-polymatroid here is called 
the
{\em contra}-polymatroid.} (function) if it holds that
\beqns
\mbox{1)} & & \sigma (\emptyset)  =  0,\\
\mbox{2)}& &\sigma (S) \le  \sigma ( T) \quad (S \subset T),\\
\mbox{3)} & & \sigma (S  \cap T) + \sigma (S \cup T) \ge  \sigma (S) + \sigma (T).
\eeqns
It is not difficult to check that $\sigma(S) = H(X_S|X_{\overline{S}})$  is a co-polymatroid
(see, Han \cite{han-cover}).
On the other hand, a set function $\rho (S)$ on $\Phi$ is called a polymatroid if it holds that
\beqns
\mbox{$1^{\prime}$)} & & \rho (\emptyset)  =  0,\\
\mbox{$2^{\prime}$)}& &\rho (S) \le  \rho ( T) \quad (S \subset T),\\
\mbox{$3^{\prime}$)} & & \rho (S  \cap T) + \rho (S \cup T) \le  \rho (S) + \rho (T).
\eeqns
It is also not difficult to check that 
for each $t\in \Psi$ the function $\rho_t (S)$ in (\ref{eq:newji-2})
is a polymatroid (cf. Han \cite{han-cover}, Meggido \cite{meggido}),
but $\rho_{\cN} (S)$ in (\ref{eq:newji-34})) is not necessarily a polymatroid.
These properties have been fully invoked in establishing the {\em matching condition} between source and channel for the special case of $|\Psi|=1$ ( cf. Han \cite{han-cover}).
In this paper too, they play a relevant role in order to argue about  the {\em separation} problem between 
distributed source coding and network coding. This problem is mentioned later in Section \ref{alter-cond} 
(cf. Remark \ref{chui:geriko1}).
\QED
}
\echui
With these preparations we will demonstrate the main result in the next section.
%
%
%
%section4.5
\section{Main Result \label{ss:HT_LD_C}}

The problem that we deal with here is not that of establishing the ``capacity region"
as usual, because the concept of ``capacity region" does not make sense for the general network with correlated sources. Instead,   we are interested in the {\em matching} problem between  the correlated source $X_{\Phi}$ and the network 
$\cN = (V,E,C)$ (transmissibility: cf. Definition \ref{teigi-newji-2}).
Under what condition is such a matching possible? This is the key problem here.
An answer to this question is just our main result to be stated here.
\bteiri\label{teiri:newji-1}
{\rm
The source $X_{\Phi}$ is transmissible over the network 
$\cN = (V,E,C)$ if and only if 
\beq\label{eq:newji-3}
H(X_S|X_{\overline{S}}) \le \rho_{\cN}(S)\quad (\emptyset \neq \forall S \subset \Phi)
\eeq
holds. \QED
}
\eteiri
%s
%\bchui
%{\rm
%%H(X_S|X_{\overline{S}})$ on the left-hand side 
%should be interpreted as the conditional entropy rate
%(cf. Cover \cite{cover}).
%On the other hand,  the channels cannot be 
%non-memoryless as will be seen in the proof below.
%}
%QED
%\echui
%
\bchui
{\rm
The case of $|\Psi| =1$ was investigated by Han \cite{han-cover}, 
and subsequently revisited by Barros and Servetto \cite{bar-serv},
while the case of $|\Phi| =1$ was investigated by Ahlswede, Cai, Li and Yeung \cite{al-yeung}.
%
%It should be noticed here  that Theorem 1 of  \cite{bar-serv} is wrong, 
%because, instead of  $\rho_{\cN}(S)$ in
%(\ref{eq:newji-3}), another quantity $\sum_{i\in S, j\in \overline{S}}c_{ij}$ 
%appears in \cite{bar-serv}.
% different from the correspondent in \cite{bar-serv} and 
%See a counter example (cf. Fig. 1), where capacities of bold edges =2 bits,
%capacity of broken edge =1 bit; $X_1, X_2$ are i.i.d. souces of bit rate =2  ($|\Phi | =2, |\Psi|=1$).
}
%
%It is easy to see that these quantities are different in the light of 
% a simple counterexample with 
%four nodes, three edges and two correlated sources ($|\Phi | =2, |\Psi|=1$), 
%which  is left to the reader.
%}
\QED
\echui

\bchui\label{chui-benpi}
{\rm
If the sources are {\em mutually independent}, (\ref{eq:newji-3}) reduces to
\[
\sum_{i \in S}H(X_i) \le 
\rho_{\cN}(S) \quad (\emptyset \neq \forall S \subset \Phi).
\]
Then, setting the rates as $R_i = H(X_i) $  we have another equivalent form:
\beq\label{eq:miyako-1}
\sum_{i \in S}R_i \le 
\rho_{\cN}(S) \quad (\emptyset \neq \forall S \subset \Phi).
\eeq
This specifies the {\em capacity region} of independent message rates in the traditional sense.
In other words, in case  the sources are  independent, the concept of capacity region makes sense.
In this case too, channel coding looks like for
{\em non-physical}  flows (as for the  case of $|\Phi| =1$, see Ahlswede, Cai,  Li and Yeung
\cite{al-yeung}; and as for the case of $|\Phi| >1$ see, e.g., Koetter and Med\'ard  
 \cite{koetter-med}, Li and Yeung \cite{li-yeung}). It should be noted  that formula (\ref{eq:miyako-1}) is {\em not} derivable by
a naive extension of the arguments as used in the case of single-source ($|\Phi|=1$), irrespective of  the comment in \cite{al-yeung}.
}
\QED
\echui

%
%\medskip 

\noindent
{\em Proof of Theorem \ref{teiri:newji-1}
}
\bigskip

{\em1. Converse part:}

\medskip

%\rm }
 Suppose that the source $X_{\Phi}$ is transmissible over the network $\cN = (V,E,C)$ 
with error probability $\lambda_{n,t} \equiv \Pr\{\hat{X}^n_{\Phi,t} \neq X^n_{\Phi}\}$
 ($t \in \Psi$) under 
encoding functions $f_{sj}, f_{ij}$ and decoding functions $g_t.$ 
It is also supposed  that $\lambda_{n,t} \to 0$ ( $n\to\infty$) with the $\tau$-capacity.

Here, the input to and the output from channel $(i.j) \in E$ may be regarded as
 random variables that were induced by the random variable $X_{\Phi}^n =
$ $(X^n_{s_1}, \cdots, X^n_{s_p})$.
In the following, we fix an element $\ssx_{\overline{S}} \in \cX^n_{\overline{S}}$, where 
$\overline{S}$ is the complement of $S$ in $\Phi$. Set
\beq\label{eq:new-old-2}
\lambda_{n,t}(\ssx_{\overline{S}}) = \Pr\{\hat{X}^n_{\Phi,t} \neq X^n_{\Phi}
 | X^n_{\overline{S}}=
\ssx_{\overline{S}}\},
\eeq
then
\beq\label{eq:new-old-3}
\lambda_{n,t}\equiv 
\Pr\{\hat{X}^n_{\Phi,t} \neq X^n_{\Phi}
\} =\sum_{\ssx_{\overline{S}}}
\Pr\{X^n_{\overline{S}} =  \ssx_{\overline{S}}\}\lambda_{n,t}(\ssx_{\overline{S}}). 
\eeq
For $\emptyset \neq S \subset \Phi $ and  $t \in \Psi$,  let $M_0$ be a minimum cut, i.e., a cut 
such that 
\beqn\label{eq:new-old-4}
\rho_t(S)&  =& \min\{c(M,V \setminus M) | S\subset M, t \in V \setminus M\}\nonumber\\
&=& c(M_0, V \setminus M_0),
\eeqn
and list all the channels $(i,j)$ such that $i\in M_0,  j \in V \setminus M_0$ as 
\beq\label{eq:new-old-5}
(i_1, j_1), \cdots, (i_r, j_r).
\eeq
Furthermore, let the input and  the output
of  channel $(i_k, j_k)$ be denoted by $Y^n_k, Z^n_k,$
respectively ($k=1,2, \cdots, r)$.
Set
\beq\label{eq:new-old-7}
Y^n =(Y^n_1,\cdots, Y^n_r), \quad Z^n =(Z^n_1,\cdots, Z^n_r).
\eeq
Since we are considering those codes $(n, (R_{ij})_{(i,j)\in E}, \delta, \vep)$ as defined
by (\ref{newji-1}) $\sim$ (\ref{newji-7}) in Section \ref{ss:HT_LD} on
an {\em acyclic} directed graph (cf. Remark \ref{oldji-1}) and hence there is no feedback, it is easy to see that
$X^n_{\Phi} \to Y^n \to Z^n \to \hat{X}_{\Phi, t}^n$ (conditioned on 
$X^n_{\overline{S}}=
\ssx_{\overline{S}}$) forms a Markov chain in this order.
Therefore, by virtue of the data processing lemma (cf. Cover and Thomas \cite{cover-thomas}), we have 
\beq\label{eq:new-old-8}
I(X^n_{\Phi}; \hat{X}^n_{\Phi, t}  | \ssx_{\overline{S}}) \le I(Y^n;Z^n |  \ssx_{\overline{S}}).
\eeq
On the other hand, noticing that $X^n_{\Phi}$ takes values in
$\cX^n_{s_1}\times \cdots \times \cX^n_{s_p}$ and applying Fano's lemma (cf. Cover and Thomas \cite{cover-thomas}), we have
\beq\label{eq:new-old-9}
H(X^n_{\Phi}|\hat{X}^n_{\Phi, t},\ssx_{\overline{S}}) \le
1 + n\lambda_{n,t}(\ssx_{\overline{S}}) \sum^p_{k=1}\log |\cX_{s_k}| \equiv r_t(n, \ssx_{\overline{S}},S).
\eeq
Hence, 
\beq\label{eq:new-old-10}
I(X^n_{\Phi}; \hat{X}^n_{\Phi, t}  | \ssx_{\overline{S}}) \ge H(X^n_{\Phi} | 
\ssx_{\overline{S}}) - r_t(n, \ssx_{\overline{S}},S).
\eeq
From (\ref{eq:new-old-8}) and (\ref{eq:new-old-10}),
\beq\label{eq:new-old-11}
H(X^n_{\Phi} | \ssx_{\overline{S}}) \le
I(Y^n;Z^n | \ssx_{\overline{S}}) + r_t(n,\ssx_{\overline{S}},S).
\eeq
On the other hand, since all the  the channels on the network  are mutually independent and
satisfy the strong converse property,  it follows by virtue of Lemma \ref{hoda-strong-cap} that
\beqn\label{eq:new-old-12}
I(Y^n;Z^n | \ssx_{\overline{S}}) & \le & \sum_{k=1}^r I(Y^n_k; Z^n_k|\ssx_{\overline{S}})\nonumber\\
& \le & n\sum_{k=1}^r \nth\max_{Y^n_k}I(Y^n_k; Z^n_k)\nonumber\\
& \le & n\sum_{k=1}^r \left(\lim_{n\to\infty} \nth\max_{Y^n_k}I(Y^n_k; Z^n_k)+\tau\right)\nonumber\\
%
%&\le& \sum_{k=1}^r
%
& =& n\sum^r_{k=1}(c_{i_k, j_k} + 2\tau )\nonumber\\
&=&n(\rho_t(S)+2r\tau)
\eeqn
for all sufficently large $n$, where the first inequality of (\ref{eq:new-old-12}) follows from 
the property that all the channels  are assumed to be mutually independent.%
\footnote{Specifically, let $U_1, \cdots,U_r; V_1,\cdots, V_r$ be random variables such that
$p(v_i|u_1, \cdots, u_r)= p(v_i|u_i)\ (i=1,\cdots, r)$ 
({\em channel independence}), then $I(U_1, \cdots,U_r; V_1,\cdots, V_r)\le 
\sum_{i=1,\cdots, k}^{r}I(U_i; V_i)$  (cf. Cover and Thomas \cite{cover-thomas}).
}

 It should
be noted here that we are now considering the $\tau$-capacity (cf. Definition \ref{teigi-newji-2}).
Thus, averaging both side of (\ref{eq:new-old-11}) and (\ref{eq:new-old-12})
with respect to $\Pr\{X^n_{\overline{S}}= \ssx_{\overline{S}}\}$, we have
\beq\label{eq:new-old-13}
\nth H(X^n_{S} | X^n_{\overline{S}}) \le \rho_t(S) + \overline{r}_t(n, S).
\eeq
where 
\[
\overline{r}_t(n, S) = \nth + \lambda_{n,t} \sum^p_{k=1}\log |\cX_{s_k}| + 2r\tau.
\]
Noting that $X^n_{\Phi}$ is stationary and ergodic and taking the limit $n\to \infty$ on  both sides of (\ref{eq:new-old-13}), it follows that
%$H(X^n_{S} | X^n_{\overline{S}}) =nH(X_{S} | X_{\overline{S}})).$
%Thus,
\beq\label{eq:new-old-14}
H(X_{S} | X_{\overline{S}}) \le \rho_t(S) + 2r\tau,
\eeq
where $H(X_{S} | X_{\overline{S}}) $ is the conditional entropy rate and we have noticed  that
  $\lambda_{n,t} \to 0$ as  $n\to\infty$. Since $\tau>0$ is arbitrarily small, we have
\beq\label{eq:new-old-15}
H(X_{S} | X_{\overline{S}}) \le \rho_t(S).
\eeq
Since $t \in \Psi$ is arbitrary, we conclude that
\[
H(X_{S} | X_{\overline{S}}) \le \rho_{\cN}(S).
\]

\bigskip

{\em 2. Direct part:}

\medskip

Suppose that inequality (\ref{eq:newji-3}) holds. It suffices to show  that for $R_{ij} =c_{ij} + \tau$ is achievable for any small $\tau>0$ (see Definitions \ref{teigi-newji-1}, \ref{teigi-newji-2}). 
To do so, we will use below the {\em random coding} argument. Before that, we need some preparation.
First, with sufficiently small $\delta>0$  in Definition \ref{teigi-newji-1}
we have
\beq\label{eq:gerilla1}
c_{ij} + \frac{\tau}{2} <
R_{ij} -\delta =  c_{ij} + \tau -\delta < c_{ij} +\tau.
\eeq
The second inequality  guarantees that, for each channel $w_{ij}$,   $\tau$-capacity  $R_{ij} =c_{ij} + \tau$  is enough, with appropriate choice of an encoder $\varphi_{ij}$ and a decoder $\psi_{ij}$,  to attain reliable reproduction of the input of the encoder $\varphi_{ij}$ (i.e., the output of $f_{ij}$ with domain size 
$2^{n(R_{ij} - \delta)}$)  at the decoder $\psi_{ij}$ with {\em maximum} decoding  error probability $\gamma_n^{(i,j)} \ge 0$ such that $\gamma_n^{(i,j)}  \to 0$ as $n \to $  $\infty$ (cf. e.g., Gallager \cite{gall}, Csisz\'ar and K\"orner \cite{csis-kor}). On the other hand,  the first inequality of (\ref{eq:gerilla1}) will be used later.
%
%\QED
%\echui
%%

In order to first evaluate the error probability 
$$\lambda_{n,t}\equiv \Pr\{\hat{X}^n_{\Phi,t} \neq X^n_{\Phi}\},$$
 let us define the error event:
%\beq\label{eq-event-1}
$$
E_n = \{\mbox{decoding errors are caused by channel coding via some} \ w_{ij}\mbox{'s}\},
%\nonumber
$$
or more formally, 
\beq\label{eq:nekom-1}
E_n = \{ h_{ij} \neq f_{ij}\mbox{\ as functions for some } (i,j)\in E\},
\eeq
where $f_{ij}$'s and $h_{ij}$'s ($(i,j) \in E$) have been specified in (\ref{newji-1}) $\sim$ (\ref{newji-7}).
Then,
\beq\label{eq-event-2}
\lambda_{n,t}= \Pr\{\overline{E}_n\}\Pr\{\hat{X}^n_{\Phi,t} \neq X^n_{\Phi} | 
\overline{E}_n\}
+ \Pr\{E_n\}\Pr\{\hat{X}^n_{\Phi,t} \neq X^n_{\Phi} | 
E_n\},
\eeq
where $\overline{E}_n$ indicates the complement of $E_n$, i.e.,
\beq\label{eq:nekom-103}
\overline{E}_n = \{ h_{ij} = f_{ij}\mbox{\ as functions for all } (i,j)\in E\}.
\eeq
Now define 
\beq\label{eq:shannon-award-1}
E_n^{(i,j)} = \{ h_{ij} \neq f_{ij}\mbox{\ as functions\}} \quad   \mbox{for}\ (i,j)\in E,
\eeq
then it is not difficult to check  that $\gamma_n^{(i,j)} = \Pr\{E_n^{(i,j)}\}$ because  $\gamma_n^{(i,j)}$ is 
the {\em maximum} decoding error probability.
Moreover, we see that 
$$E_n = \cup_{(i,j)\in E} E_n^{(i,j)} \ \mbox{(disjoint union)}.$$
Therefore,   
\beqn\label{eq-event-3}
\lambda_{n,t} & \le&  \Pr\{\hat{X}^n_{\Phi,t} \neq X^n_{\Phi} | 
\overline{E}_n\} +  \Pr\{E_n\}\nonumber\\
&= &  \Pr\{\hat{X}^n_{\Phi,t} \neq X^n_{\Phi} | 
\overline{E}_n\} + \sum_{(i,j)\in E}\Pr\{E_n^{(i,j)}\}\nonumber\\
&= &  \Pr\{\hat{X}^n_{\Phi,t} \neq X^n_{\Phi} | 
\overline{E}_n\} + \sum_{(i,j)\in E}\gamma_n^{(i,j)}\nonumber\\
&\le & \Pr\{\hat{X}^n_{\Phi,t} \neq X^n_{\Phi} | 
\overline{E}_n\} + |E|\gamma_n
\eeqn
with $\gamma_n =\max_{(i,j)\in E}\gamma_n^{(i,j)}$, where   the first equality comes from the fact that all component channels are {\em independent}.
 It is obvious  that $\gamma_n \to 0$ as $n \to\infty$.%The reason for the appearance of $2|E|\gamma_n$ instead of 
%$|E|\gamma_n$ is that we consider below channel coding 
%for two different sources $\ssx_{\Phi} \neq
%\ssx^{\prime}_{\Phi}$.

Thus, in order to demonstrate $\lambda_{n,t}\to 0$, it suffices to show that 
\beq\label{eq-event-5}
\beta_{n,t}\equiv \Pr\{\hat{X}^n_{\Phi,t} \neq X^n_{\Phi} | 
\overline{E}_n\}
\to 0 \quad (n\to\infty),
\eeq
which means that we may assume throughout in the sequel that all the channels in the network are regarded as {\em noiseless} (i.e., the {\em identity} mappings).
Accordingly, then, $h_{ij}\equiv \psi_{ij}\circ w_{ij}\circ \varphi_{ij}\circ f_{ij}$ reduces to 
$h_{ij} = f_{ij}$ 
with domain size $2^{n(R_{ij}-\delta)}$, and consequently $\tilde{h}_{ij} = \tilde{f}_{ij}$, where 
$\tilde{f}_{ij}$ denotes the value of $f_{ij}$ as a function of $\ssx_{\Phi};
$
similarly for $\tilde{h}_{ij}$.  Thus, we can {\em separate}  channel coding from network coding.
Hereafter, for this reason, we use only the notation $f_{ij}, \tilde{f}_{ij}$ instead of
$h_{ij}, \tilde{h}_{ij}$. 

Let us now return to show, in view of Definition \ref{teigi-newji-2}, that 
$(c_{ij} + \tau)_{(i,j)\in E}$ is achievable for any snall $\tau >0$. To do so, we 
invoke the {\em random coding} argument: for each 
$$z\in  \prod_{k:(k,i)\in E} [1, 2^{n(R_{ki}-\delta)}],$$
 make  $f_{ij}(z)$ take values {\em uniformly} and {\em independently}  in $[1, 2^{n(R_{ij} - \delta)}]$
 (cf. (\ref{newji-3})).
%
%construct the sought-for encoding functions as follows. 
%where we set $\vep >0$ is arbitrarily small.
First, define the associated random variables, as functions of $\ssx_{\Phi} \in \cX^n_{\Phi}$,  such that
\[
z_s(\ssx_{\Phi}) = \ssx_s\quad (s \in \Phi),
\]
\[
z_j(\ssx_{\Phi}) = (\tilde{f}_{kj}(\ssx_{\Phi}))_{(k,j)\in E} \quad (j\not\in \Phi).
\]
It is evident that $z_j(\ssx_{\Phi})$'s thus defined carry on  all the information received at node $j$ during the coding process.

In the sequel we use the following notation: fix an $\ssx_{\Phi} \in \cX^n_{\Phi}$ and
decompose it as  $\ssx_{\Phi} = ( \ssx_{S},  \ssx_{\overline{S}})$ where 
($\emptyset \neq S \subset \Phi)$. We indecate by $\ssx^{\prime}_{\Phi [S]}$
an $\ssx^{\prime}_{\Phi} = ( \ssx^{\prime}_{S},  \ssx^{\prime}_{\overline{S}})$ 
such that $ \ssx^{\prime}_{S} \neq \ssx_{S} $,
$\ssx^{\prime}_{\overline{S}} = \ssx_{\overline{S}}$,  where 
$\ssx^{\prime}_{S} \neq \ssx_S $ means  componentwise unequality, i.e., 
$\ssx^{\prime}_s \neq \ssx_{s}$ for all $s\in S$. It should be remarked here that 
two  distinct  sequences $\ssx^{\prime}_{\Phi [S]}\neq \ssx_{\Phi}$ are
 indistinguishable at the decoder $t\in \Psi$ if and only if $z_t(\ssx_{\Phi}) =
  z_t(\ssx^{\prime}_{\Phi [S]})$.
  The proof to be stated below is basically along in the same spirit as that of 
  Ahlswede, Cai, Li and Yeung \cite{al-yeung}, although we need here to invoke the  joint
  typicality argument as well as  subtle arguments on the classification of error patterns.
  
  Let us now evaluate the probability of decoding error under the encoding scheme
  as specified in Section \ref{ss:HT_LD}.$C$.
  We first fix a typical sequence $\ssx_{\Phi}\in T_{\lambda}(X_{\Phi})$,
   and for $t \in \Psi$ and $\emptyset \neq S \subset \Phi$, define
  % \[
\beq\label{newji-array-1}
 F_{S,t}(\ssx_{\Phi}) =\left\{
\begin{array}{ll}
 1   & \mbox{if there exists some}\  \ssx^{\prime}_{\Phi [S]} \neq \ssx_{\Phi}\  \mbox{such that}\\
 &
\ssx^{\prime}_{S}\  \mbox{is jointly typical with}\  \ssx_{\overline{S}}\  \mbox{and}
\ z_t(\ssx_{\Phi}) =
  z_t(\ssx^{\prime}_{\Phi [S]}),\nonumber\\
  0 &  \mbox{otherwise.}
  \end{array}\right.
 \eeq 
 Furthermore, set
 \beq\label{er:newji-7} 
 F(\ssx_{\Phi}) = \max_{\emptyset \neq S \subset \Phi, t\in \Psi} F_{S,t}(\ssx_{\Phi}), 
 \eeq
 where we notice that $ F(\ssx_{\Phi}) =$  1 if and only if  $\ssx_{\Phi}$
 cannot be uniquely recovered by at least one sink node $t\in \Psi$.%\] 

Here, for any node $i\in V$ let ${\cal D}_i$ denote the set of all the starting nodes of the longest directed paths
 ending at node $i$, and set
 %Without loss of generality we may assume
%\[
%{\cal D} \equiv \cup_{i\in V}{\cal D}_i \subset \Phi.
%\] 
%and set
 \[
 V_0 = \{i\in V| \Phi \cap \cD_i \neq \emptyset \}\mbox{\ and}\   V_1 \equiv V \setminus V_0.
 \]
Furthermore, we consider any $\ssx^{\prime}_{\Phi [S]} \neq \ssx_{\Phi}$ and define
 \beqn\label{eq:jiu-1}
 B_0 & =& \{ i\in V_0 |  z_i(\ssx_{\Phi}) \neq
  z_i(\ssx^{\prime}_{\Phi [S]}) \},\\
   B_1 & =& \{ i\in V_0 |  z_i(\ssx_{\Phi}) =
  z_i(\ssx^{\prime}_{\Phi [S]}) \},
  \eeqn
  where $B_0$ is the set of nodes $i$ at which two sources
  $\ssx_{\Phi} $ and $
  \ssx^{\prime}_{\Phi [S]}$ are distinguishable, and $ B_1\cup V_1$ is the set of nodes  $i$ at which 
  $\ssx_{\Phi} $ and $\ssx^{\prime}_{\Phi [S]}$ are indistinguishable.
  It is obvious that $S \subset  B_0 \subset V_0$, $\overline{S} \subset B_1$
  and $B_1\cup V_1=V\setminus B_0$.   %

 % We consider an arbitrary partition of $V_1$ such that 
 %$V_1 =C_0\cup C_1\ (C_0\cap C_1 = \emptyset)$, and 
%Set $N_0 = B_0\, \  N_1 = B_1\cup V_1$.
%  $\overline{S} \subset B_1$ by convention.
 % 
  Now let us fix any $\ssx_{\Phi}$ and suppose that $z_t(\ssx_{\Phi}) =
  z_t(\ssx^{\prime}_{\Phi [S]})$, which implies that  
   $t \in B_1$.
Then,  we see that $B_0 = N$ for some $N \subset V$ such that $S \subset N$ and $t\not\in N$, that is, $N$ is a fixed cut between $S$ and $t$. 
%where  $N$ is meant to be  a {\em deterministic} cut.
  %
  %
  %
%Let us use here the {\em random coding} argument in order to have a ``good" 
%encoding function for every $f_{ij}$: 
%For each 
%$$z\in  \prod_{k:(k,i)\in E} [1, 2^{n(R_{ki}-\delta)}],$$
% make  $f_{ij}(z)$ take values {\em uniformly} and {\em independently} 
% in $[1, 2^{n(R_{ij} - \delta)}]$
% (cf. (\ref{newji-3})).
  Then, for $i \in B_0 $ and $(i,j)\in E$,
  \beqn\label{ben-kai-1}
 & &  \Pr\{\tilde{f}_{ij}(\ssx_{\Phi}) = \tilde{f}_{ij}(\ssx^{\prime}_{\Phi [S]}) |
  z_i (\ssx_{\Phi}) \neq   z_i (\ssx^{\prime}_{\Phi [S]})\}\nonumber\\
 & & = 2^{-n(R_{ij} - \delta)}\nonumber\\
 & &  \le 2^{-n(c_{ij}+\frac{\tau}{2})},
  \eeqn
  where we have used the first inequality in (\ref{eq:gerilla1}).
%  while if, $i \in V_1$,
%\beq\label{eq:maeto1}
% \Pr\{\tilde{f}_{ij}(\ssx_{\Phi}) = \tilde{f}_{ij}(\ssx^{\prime}_{\Phi [S]}) |
 % z_i (\ssx_{\Phi}) \neq  z_i (\ssx^{\prime}_{\Phi [S]})\} = 1
 %
 %\eeq 
 % by convention.
 Notice here that $B_0, B_1$ are random sets under the random coding for $f_{ij}$'s.
Therefore,  
  \beqn\label{ben-kai-13}
\lefteqn{ \Pr\{B_0 = N\}}\nonumber\\
 & =& \Pr\{B_0 = N, B_0 \supset N \}\nonumber\\
 & =& \Pr\{B_0 = N|B_0 \supset N \}\Pr\{B_0\supset N\}\nonumber\\
 %&=& \Pr\{B_0 = N | B_0 \supset N \}\Pr\{ B_0 \supset N\}\nonumber\\
 &\le & \Pr\{B_0 = N | B_0 \supset N \}\nonumber\\
 &\le & \prod_{(i,j)\in E_{N}}
  \Pr\{\tilde{f}_{ij}(\ssx_{\Phi}) = \tilde{f}_{ij}(\ssx^{\prime}_{\Phi [S]})
  | z_i (\ssx_{\Phi}) \neq   z_i (\ssx^{\prime}_{\Phi [S]}\}\nonumber\\
  &\le & \prod_{(i,j)\in E_N}2^{-n(c_{ij}+\frac{\tau}{2})}\nonumber\\
  &\le & 2^{-n(\sum_{(i,j)\in E_N}c_{ij}+\frac{\tau}{2})},
  \eeqn
  where $E_N= \{(i,j)\in E | i \in N, j \in V \setminus N\}$.
%where we have used the first inequality of (\ref) and 
%we have used the convention that,
%for $i\in  C_0$,
%\[
 % \Pr\{\tilde{f}_{ij}(\ssx_{\Phi}) = \tilde{f}_{ij}(\ssx^{\prime}_{\Phi [S]})
 % | z_i (\ssx_{\Phi}) \neq   z_i (\ssx^{\prime}_{\Phi [S]}\}=1.
 %
 % \]
Furthermore, 
  \beqn\label{ben-kai-13rt}
 \sum_{(i,j)\in E_N} c_{ij} &\ge &  \min_{N: S\subset N, t\not\in N}
 c_{ij}\nonumber\\
%& =&  \sum_{(i,j)\in E_{N^{\prime}}} c_{ij} \\
%& \ge & \min_{N: S\subset N, t\not\in N}
% c_{ij}\nonumber\\
 & =&\rho_t(S),
  \eeqn
  %
 %
% rime}}}2^{-n(c_{ij}+\frac{\tau}{2})}
where $\rho_t(S)$ was specified in Section \ref{ss:HT_LD}. 
%
%
%

%are defined above 
In conclusion, it follows from (\ref{ben-kai-13}) and (\ref{ben-kai-13rt}) that,
for any fixed cut $N$ separating $S$ and $t$, 
\beq\label{renren-1}
\Pr\{B_0 = N\} \le 2^{-n(\rho_t(S) + \frac{\tau}{2})},
\eeq
so that 
 \beqn\label{ben-kai-5}
& &\Pr\{z_t(\ssx_{\Phi}) =
  z_t(\ssx^{\prime}_{\Phi [S]})\}\nonumber\\
  & & = \Pr\{B_0 =N \mbox{\ for some cut $N$  between $S$ and $t$}\}\nonumber\\
  & & \le 2^{|V|}2^{-n(\rho_t(S) + \frac{\tau}{2})}.
  \eeqn
  On the other hand, as is seen from the definition of $F_{S,t}(\ssx_{\Phi})$ in (\ref{newji-array-1}),
  condition $F_{S,t}(\ssx_{\Phi})=1$ is equivalent to the statement 
 ``$z_t(\ssx_{\Phi}) = 
 z_t(\ssx^{\prime}_{\Phi [S]})$ for some $\ssx^{\prime}_{\Phi [S]} \neq \ssx_{\Phi}$
 such that $\ssx^{\prime}_{S}$ is jointly typical with $\ssx_{\overline{S}}.$" 
As a consequence, by virtue of Lemma \ref{nu-hodai} and (\ref{ben-kai-5}), we obtain
\beqn\label{ben-kai-9}
\Pr\{F_{S,t}(\ssx_{\Phi})=1\} &\le & 
2^{n(H(X_S|X_{\overline{S}} )+ 2\lambda)}
\Pr\{z_t(\ssx_{\Phi}) =
  z_t(\ssx^{\prime}_{\Phi [S]})\}\nonumber\\
  & \le & 2^{|V|}2^{n(H(X_S|X_{\overline{S}}) + 2\lambda -\rho_t(S)-\frac{\tau}{2})}\nonumber\\
  &\le & 
  2^{|V|}2^{-n(\rho_t(S)-H(X_S|X_{\overline{S}}) +\frac{\tau}{4})},
  \eeqn
where we have chosen $\lambda = \frac{3\tau}{8}$, since $\lambda >0$ can be arbitrarily small.
Then, in view of (\ref{er:newji-7}), it follows that
\beqn\label{ben-kai-10}
\lefteqn{\Pr\{F(\ssx_{\Phi})=1\}}\nonumber\\ & = & \Pr\{\max_{\emptyset \neq S \subset \Phi, t\in \Psi} F_{S,t}(\ssx_{\Phi}) =1\}\nonumber\\
&\le & \sum_{\emptyset \neq S \subset \Phi, t\in \Psi} \Pr\{F_{S,t}(\ssx_{\Phi})=1\}\nonumber\\
& \le &\sum_{\emptyset \neq S \subset \Phi, t\in \Psi} 
2^{|V|}
2^{-n(\rho_t(S)-H(X_S|X_{\overline{S}}) +\frac{\tau}{4})},
%
%\end{eqnarray}
%
\eeqn
  which together with condition (\ref{eq:newji-3}) yields
  \beq\label{tesun-1}
  E(F(\ssx_{\Phi})) = \Pr\{F(\ssx_{\Phi})=1\} \le 2^{-cn}\quad (\mbox{for}\ \ssx_{\Phi}\in T_{\lambda}
  (X_{\Phi}))
  \eeq
for all sufficiently large $n\ge n_0,$
where  $c=\frac{\tau}{8}$  and $E$ denotes the expectation due to random coding.

Finally, in order to show the existence of a {\em deterministic} code to attain the transmissibility over network
$\cN = (V, E,  C)$, set 
\[
G_n(\ssx_{\Phi}) = E(F(\ssx_{\Phi}))\  \mbox{for}\ \ssx_{\Phi}  \in 
 T_{\lambda}(X_{\Phi}),
\]
and set $F(\ssx_{\Phi})=1$ for $\ssx_{\Phi}  \not\in 
 T_{\lambda}(X_{\Phi})$, then, again by Lemma \ref{nu-hodai},
\beqn\label{ben-kai-12}
\sum_{\ssx_{\Phi} \in \cX^n_{\Phi}}p(\ssx_{\Phi})G_n(\ssx_{\Phi}) & = & 
\sum_{\ssx_{\Phi}\in T_{\lambda}(X_{\Phi})}p(\ssx_{\Phi})G_n(\ssx_{\Phi})  +
\sum_{\ssx_{\Phi}\not\in T_{\lambda}(X_{\Phi})}p(\ssx_{\Phi})G_n(\ssx_{\Phi})\}\nonumber\\
&\le& \sum_{\ssx_{\Phi}\in T_{\lambda}(X_{\Phi})}p(\ssx_{\Phi})G_n(\ssx_{\Phi})  +
\Pr\{X^n_{\Phi}\not\in T_{\lambda}(X_{\Phi})\}\nonumber\\
&\le& \sum_{\ssx_{\Phi}\in T_{\lambda}(X_{\Phi})}p(\ssx_{\Phi})2^{-cn}  +\lambda\nonumber\\
&\le& 2^{-cn} +\lambda.
 \eeqn
On the other hand, the left-hand side of (\ref{ben-kai-12}) is rewritten as
\beqns
& &\sum_{\ssx_{\Phi} \in \cX^n_{\Phi}}p(\ssx_{\Phi})G_n(\ssx_{\Phi})\\
& & = E(\sum_{\ssx_{\Phi} \in \cX^n_{\Phi}}p(\ssx_{\Phi})F(\ssx_{\Phi}))\\
& & =E( \mbox{ the probability of decoding error via network}\  \cN = (V,E,C)).
\eeqns
Thus, we have shown that there exists at least one deterministic code with probability of decoding error at most $2^{-cn} +\lambda$.
\section{Examples\label{ss:geri-ex1}}

In this section we show two examples of Theorem \ref{teiri:newji-1} with 
$\Phi =\{s_1. s_2\}$ and $\Psi =\{t_1. t_2\}$.

\medskip

\noindent
{\em Example 1}.\ Consider the network as in Fig.1(called the {\em butterfly}) where all the solid edges have capacity 1
and the independent sources $X_1, X_2$ are binary and uniformly distributed (cited from Yan, Yang and Zhang \cite{zhang}). The capacity function of this network is computed as follows:
\beqns
\rho_{t_1}(\{s_2\}) & = & \rho_{t_2}(\{s_1\}) = 1,\\
\rho_{t_1}(\{s_1\}) & = & \rho_{t_2}(\{s_2\}) = 2,\\
\rho_{t_1}(\{s_1,s_2\}) & = & \rho_{t_2}(\{s_1,s_2\}) = 2;
\eeqns
\beqns
\rho_{{\cal N}}(\{s_1\}) & = & \min (\rho_{t_1}(\{s_1\}),\rho_{t_2}(\{s_1\}))=1,\\
\rho_{{\cal N}}(\{s_2\}) & = & \min (\rho_{t_1}(\{s_2\}),\rho_{t_2}(\{s_2\}))=1,\\
%\rho_{{\cal N}} & = & \min (\rho_{t_2}(\{s_1),\rho_{t_1}(\{s_1))=1,\\
\rho_{{\cal N}}(\{s_1,s_2\}) & = & \min (\rho_{t_1}(\{s_1,s_2\}),\rho_{t_2}(\{s_1,s_2\}))=2.
\eeqns
On the other hand, 
\beqns
H(X_1|X_2) & = & H(X_1) =1,\\
H(X_2|X_1) & = & H(X_2) =1,\\
H(X_1X_2) & = & H(X_1) + H(X_2) =2.
\eeqns

\begin{figure}[htbp]
   \centering
   \includegraphics[bb= 0 0 200 320, scale=0.4]{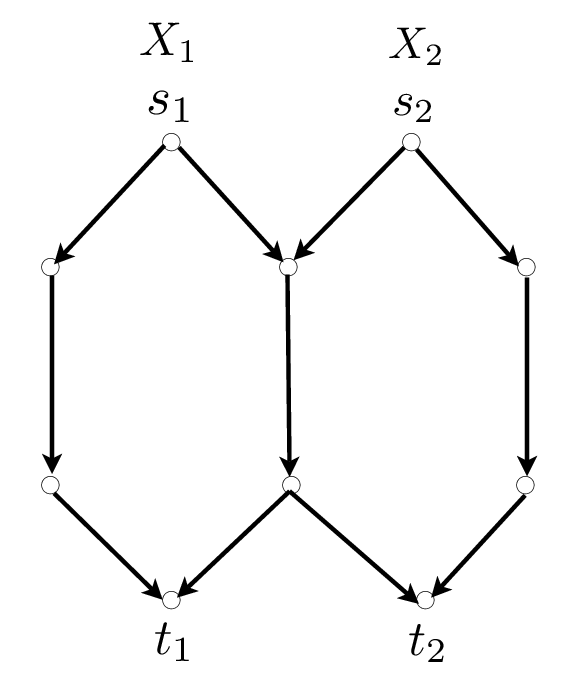} 
   \caption{Example 1}
\end{figure}

Therefore, condition (\ref{eq:newji-3}) in Theorem \ref{teiri:newji-1} is satisfied with equality, so that the sourse is transmissible over the network. Then, how to attain this transmissibility? That is depicted in Fig.2 where $\oplus$ denotes the exclusive OR. Fig. 3 depicts the corresponding  capacity region,
which is within the framework of the previous work (e.g., see Ahlswede {\em et al.} \cite{al-yeung}).

\begin{figure}[htbp]
   \centering
   \includegraphics[bb= 0 0 200 320, scale=0.4]{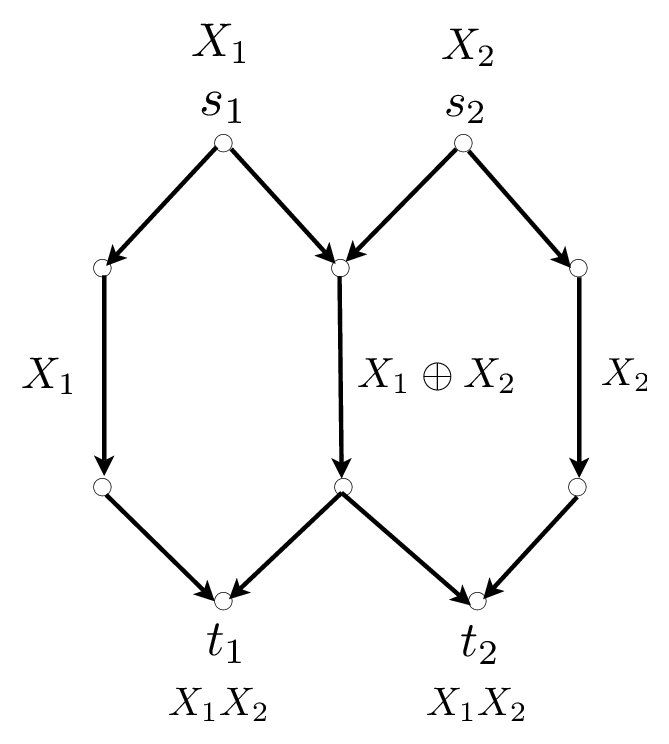} 
   \caption{Coding for Example 1}
\end{figure}

\begin{figure}[htbp]
   \centering
   \includegraphics[bb= 0 0 200 320, scale=0.5]{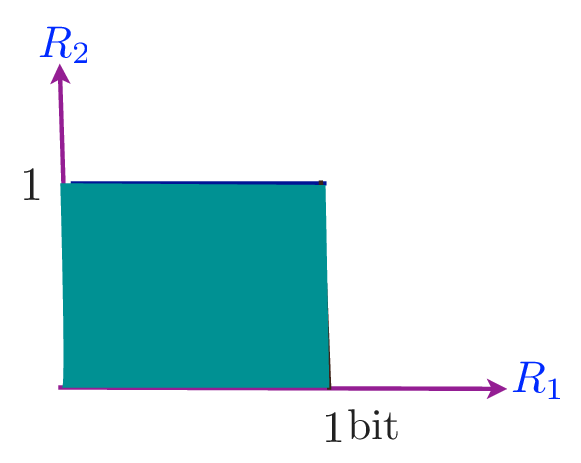} 
   \caption{Capacity region for Example 1}
\end{figure}

\medskip

\noindent
{\em Example 2}. \ Consider the network with {\em noisy} channels as in Fig.4 where the solid edges have capacity 1 and the broken edges have capacity $h(p)<1 $. Here, $h(p)$ ($0< p < \frac{1}{2}$) is the binary entropy defined by
$ h(p) = -p\log_2 p -(1-p)\log_2 (1-p).$ The source $(X_1, X_2)$ 
generated at the nodes $s_1, s_2$ is the binary symmetric source with crossover probability $p$,
i.e.,
\[
\Pr\{X_1=1\} = \Pr\{X_1=0\} =\Pr\{X_2=1\}=\Pr\{X_2=0\}=\frac{1}{2},
\]
\[
\Pr\{X_2=1|X_1=0\} =\Pr\{X_2=0|X_1=1\}=p.
\]
 Notice that 
$X_1, X_2$  are not independent.
The capacity function of this network is computed as follows:
\beqns
\rho_{t_1}(\{s_2\}) & = & \rho_{t_2}(\{s_1\}) = h(p),\\
\rho_{t_1}(\{s_1\}) & = & \rho_{t_2}(\{s_2\}) =1+h(p),\\
\rho_{t_1}(\{s_1,s_2\}) & = & \rho_{t_2}(\{s_1,s_2\}) = 2;
\eeqns
\beqns
\rho_{{\cal N}}(\{s_2\}) & = & \min (\rho_{t_1}(\{s_2\}),\rho_{t_2}(\{s_2\}))=h(p),\\
\rho_{{\cal N}}(\{s_1\}) & = & \min (\rho_{t_1}(\{s_1\}),\rho_{t_2}(\{s_1\}))=h(p),\\
%\rho_{{\cal N}} & = & \min (\rho_{t_2}(s_1),\rho_{t_1}(s_1))=1,\\
\rho_{{\cal N}}(\{s_1,s_2\}) & = & \min (\rho_{t_1}(\{s_1,s_2\}),\rho_{t_2}(\{s_1,s_2\}))=2.
\eeqns
On the other hand, 
\beqns
H(X_1|X_2) & = &h(p),\\
H(X_2|X_1) & = &h(p),\\
H(X_1X_2) & = & 1+h(p).
\eeqns

\begin{figure}[htbp]
   \centering
   \includegraphics[bb= 0 0 200 320, scale=0.4]{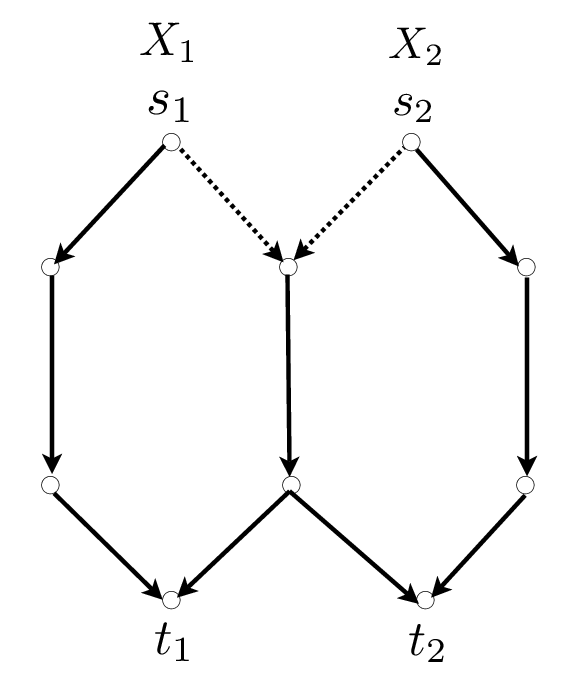} 
   \caption{ Example 2}
\end{figure}

\begin{figure}[htbp]
   \centering
   \includegraphics[bb= 0 0 200 320, scale=0.4]{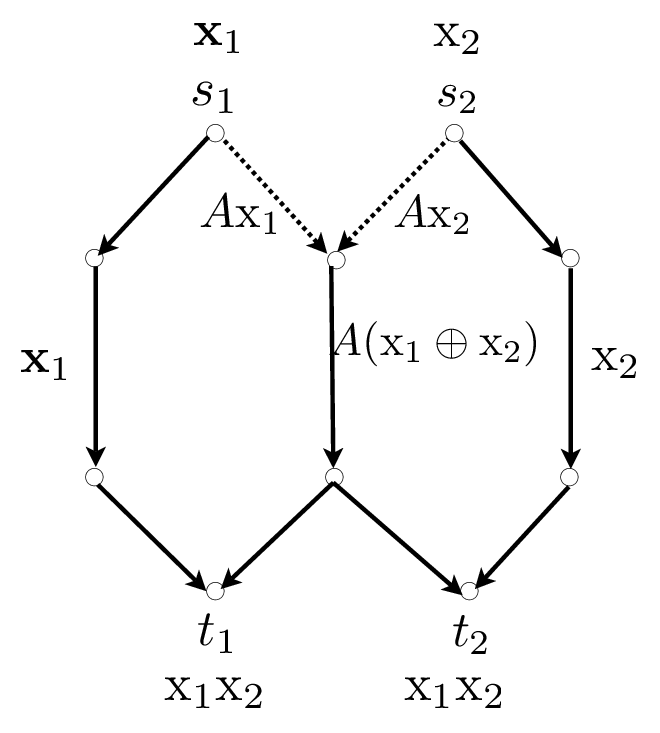} 
   \caption{Coding for  Example 2}
\end{figure}

Therefore, condition (\ref{eq:newji-3}) in Theorem \ref{teiri:newji-1} is satisfied with strict inequality, so that the source is transmissible over the network. Then, how to attain this transmissibility? That is depicted in Fig.5 where  $\ssx_1, \ssx_2$ are $n$ independent copies of $X_1, X_2$, respectively,
and $A$ is an $m\times n$ matrix ($m=nh(p)<n)$.
Notice that the entropy  of $\ssx_1\oplus \ssx_2$ (componentwise exclusive OR) is  $nh(p)$ bits and hence it is possible to recover $\ssx_1\oplus \ssx_2$ from $A(\ssx_1\oplus \ssx_2)$ (of length $m =nh(p)$) with asymtoticaly negligible probability of decoding error,  provided that $A$ is appropriately chosen 
(see K\"orner and Marton \cite{korn-mart}). It should be remarked that this example cannot be justified by the previous works such as Ho {\em et al.} \cite{koetter}, Ho {\em et al.} \cite{koetter-1}, and
Ramamoorthy {\em et al.} \cite{effros}, because all of them assume  {\em noiseless} channels with 
capacity of  {\em one bit}, i.e.,
this example is outside the previous framework. 

\section{Alternative Transmissibility Condition}\label{alter-cond}
In this section we demonstrate an alternative transmissibility condition equivalent to the 
necessary and sufficient condition  (\ref{eq:newji-3}) given 
in Theorem \ref{teiri:newji-1}. 

To do so, for each $t\in \Psi $ we define the polyhedron $\cC_t$ as the set of 
all nonnegative rates
$(R_s; s \in \Phi)$ such that 
\beq\label{eq:efros-san}
\sum_{i \in S}R_i \le 
\rho_{t}(S) \quad (\emptyset \neq \forall S \subset \Phi),
\eeq
where $\rho_{t}(S) $ is the capacity function as defined in (\ref{eq:newji-2}) of Section \ref{ss:HT_LD}.
Moreover, define the polyhedron $\cR_{\mbox{{\scriptsize SW}}}$ as the set of all nonnegative
rates
$(R_s; s \in \Phi)$ such that 
\beq\label{eq:efros-san-2}
H(X_S|X_{\overline{S}}) \le \sum_{i \in S}R_i  \quad (\emptyset \neq \forall S \subset \Phi),
\eeq
where $H(X_S|X_{\overline{S}})$ is the conditional entropy rate as defined in Section \ref{ss:HT_LD}.
Then, we have the following theorem on the transmissibility over the network $\cN=(V, E, C)$.
\bteiri\label{teiri:odoroi-te}
{\rm The following two statements are equivalent:
\beqn
 1)& & H(X_S|X_{\overline{S}}) \le \rho_{\cN}(S)\quad (\emptyset \neq \forall S \subset \Phi)\label{eq:teruo-1},\\
 2) & & \cR_{\mbox{{\scriptsize SW}}}\cap \cC_t \neq \emptyset
\quad (\forall t\in \Psi).\label{eq:teruo-2}
\eeqn
}
\eteiri  
In order to prove Theorem \ref{teiri:odoroi-te} we need the following lemma:
\bhodai [{\rm Han \cite{han-cover}}]\label{hodai:kopit-1} %\mbox{}
{\rm 
Let $\sigma(S)$, $\rho(S)$ be a co-polymatroid and a polymatroid, respectively, 
as defined in Remark \ref{chui-aho1}. Then,  the necessary and sufficient condition for the 
existence of some nonnegative rates  $(R_s; s \in \Phi)$ such that
 \beq\label{eq:mannax-3}
 \sigma (S) \le \sum_{i \in S} R_i   \le   \rho (S)   \quad (\emptyset \neq \forall S \subset \Phi)
\eeq
is that
  \beq\label{eq:mannax-4}
\sigma (S) \le  \rho (S)\quad (\emptyset \neq \forall S \subset \Phi).
\eeq
\QED
}
\ehodai

\medskip

\noindent
{\em  Proof of Theorem \ref{teiri:odoroi-te} }:

Suppose that (\ref{eq:teruo-1} ) holds, then, in view of (\ref{eq:newji-34}), this implies
 \beq\label{eq:mannax-6}
 H(X_S|X_{\overline{S}}) \le \rho_{t}(S)\quad (\forall t\in \Psi, \emptyset \neq \forall S \subset \Phi).
 \eeq
 Since, as was pointed out in Remark \ref{chui-aho1}, $\sigma (S) = H(X_S|X_{\overline{S}})$ and
 $\rho (S) = \rho_t (S)$ are a co-polymatroid and a polymatroid, respectively, 
 application of Lemma \ref{hodai:kopit-1} ensures the existence of some nonnegative rates 
  $(R_s; s \in \Phi)$ such that
 \beq\label{eq:mannax-7}
  H(X_S|X_{\overline{S}}) \le \sum_{i \in S} R_i \le \rho_t (S)\quad (\forall t\in \Psi, 
  \emptyset \neq \forall S \subset \Phi), 
\eeq
  which is nothing but (\ref{eq:teruo-2}).
  
  Next, suppose that (\ref{eq:teruo-2}) holds. This implies  (\ref{eq:mannax-7}), which in turn implies (\ref{eq:mannax-6}), i.e., (\ref{eq:teruo-1}) holds. 
  \QED
\bchui\label{chui:toutou-2}
{\rm  
The necessary and sufficient condition of the form (\ref{eq:teruo-2}) appears ({\em without} the proof) in Ramamoorthy, Jain, Chou
 and Effros \cite{effros} with $|\Phi| =2, |\Psi|=2$, which they call the {\em feasibility}. They attribute the sufficiency part simply to Ho, M\'edard, Effros and Koetter \cite{koetter} with $|\Phi| =2, |\Psi|=1$ (also, cf. 
Ho, M\'edard,  Koetter, Karger, Effros, Shi, and Leong \cite{koetter-1} with $|\Phi| =2, |\Psi|=1$),
 while attributing  the necessity part to Han \cite{han-cover}, Barros and  Servetto \cite{bar-serv}.
 However,  notice that 
 %it is assumed  in \cite{koetter} and  \cite{koetter-1} 
 %(\cite{koetter} is included in \cite{koetter-1}) that 
% the paths from source nodes to sink nodes are  all of the same length $L$;  and 
 %
  all the arguments in \cite{koetter}, \cite{koetter-1} 
  (\cite{koetter} is included in \cite{koetter-1})
  can be validated  only  within  the class of stationary memoryless sources of {\em  integer} bit rates and 
 {\em error-free} channels (i.e., the {\em identity} mappings) all with {\em one bit} capacity
   (this restriction is needed to invoke ``Menger's theorem" in graph theory); 
% (allowing multiple edges with {\em one bit} capacity); %furthermore, there, the very stringent restriction is imposed that all the paths 
 %connecting source nodes and sink nodes are all of the same length $L$. 
 while the present  paper, without such severe restrictions,  treats  ``general" acyclic networks, allowing  for general correlated stationary ergodic sources as well as general statistically  independent channels with each satisfying the strong converse property (cf. Lemma \ref{hoda-strong-cap}). Moreover, as long as we are concerned also  with %general networks also with 
 {\em noisy} channels, the way of approaching the problem as in \cite{koetter}, \cite{koetter-1} 
does {\em not} work as well, because in this noisy  case we have to cope with two kinds of error probabilities, 
 one due to error probabilities for source coding and the other due to error probabilities for network coding (i.e., channel coding); thus in the noisy channel case or in the noiseless channel case with {\em non-integer} capacities
 and/or i.i.d. sources of  {\em non-integer} bit rates, \cite{effros} cannot attribute the sufficiency part 
 of (\ref{eq:teruo-2}) to \cite{koetter}, \cite{koetter-1}. 
 
 It should be noted here also that   \cite{koetter} and \cite{koetter-1}, though demonstrating relevant error exponents (the {\em direct} part),  do not have the {\em converse} part.
 \QED
}
\echui 
\bchui [{\it Separation}]\label{chui:geriko1}
{\rm
Here, the term of {\em separation} is used to mean separation of distributed source coding and network coding with {\em independent} sources. 
%We here prefer the latter standpoint.
%
Theorem \ref{teiri:newji-1} does not immediately guarantee {\em separation} in this sense. 
%
%because $\rho_{\cN}(S)$ is not necessarily a polymatroid as mentioned in Remark \ref{chui-aho1}. 
%
However, when $\rho_{\cN}(S)$ is, for example,   a polymatroid as mentioned in Remark \ref{chui-aho1}, 
{\em separation} in this sense is ensured, because in this case it is guaranteed by Lemma \ref{hodai:kopit-1}
 that there exist some nonnegative rates $R_i$ $ (i\in \Phi)$ such that
\beq\label{look-like-1}
H(X_S|X_{\overline{S}}) \le \sum_{i \in S}R_i \le 
\rho_{\cN}(S) \quad (\emptyset \neq \forall S \subset \Phi).
\eeq	
Then, the first inequality ensures {\em reliable} distributed source coding by virtue of  the 
theorem of Slepian and Wolf (cf. Cover \cite{cover}), while the second
inequality ensures {\em reliable} network  coding, that looks like for {\em non-physical}  flows,   with {\em independent} distributed sources of rates $R_i$ ($i \in \Phi$; see Remark \ref{chui-benpi}). 
%This kind of observation has previously been  found  in \cite{effros}.
Furthermore,  in the particular case of $|\Psi| =1$,  the capacity function $\rho_{\cN}(S)$ is
always  a polymatroid, so {\em separation} holds, where
network coding looks like for {\em physical}  flows (cf. Han \cite{han-cover}, Meggido \cite{meggido}, and Ramamoorthy,  Jain, Chou and Effros \cite{effros}). Then, it would be  natural to ask the question whether
separability in this sense implies polymatroidal property. 
%An answer (given in \cite{effros}) to this question is 
In this connection, \cite{effros} claims
 that, in the case with
$|\Phi | = |\Psi | = 2$ and with {\em rational} capacities as well as sources of {\em integer}
 bit rates, ``{\em separation}"  always holds, irrespective of the polymatroidal property, while
 in the case of $|\Phi|>2$ or $|\Psi|>2$ no
 conclusive claim is not made. On the other hand, we notice here that condition (\ref{look-like-1}) is  actually {\em sufficient}
 for separability despite  the non-polymatroid property of $\rho_{\cN}(S)$.
 %although 
 %
 Condition (\ref{look-like-1}) is equivalently written as  
 \beq\label{eq:equiq1}
 \displaystyle{
 \cR_{\mbox{{\scriptsize SW}}}\cap 
\left(\bigcap_{t\in \Psi} \cC_t\right) \neq \emptyset
}
 \eeq
 for any general network $\cN$. Moreover, in view of Remark \ref{chui-benpi}, it  is not difficult to check that
 (\ref{eq:equiq1}) is also necessary.
 Thus, 
our conclusion is  that, in general,  condition (\ref{eq:equiq1}) is not only {\em sufficient} but 
also {\em necessary} for separability.%
 %It is not difficult to check that condition (\ref{eq:equiq1}) is also {\em necessary} in general.
% it is currently
% not quite sure.
}
\QED
\echui

\bchui\label{chui:toutou-A3-1}
{\rm 
It is possible also to consider  network coding with {\em cost}. In this regard the reader may refer to, e.g., 
Han \cite{han-cover}, Ramamoorthy \cite{rama}, Lee {\em et al.} \cite{lee}.
\QED
}
\echui
\bchui\label{chui:toutou-A3}
{\rm 
So far we have focused on  the  case where the channels of a network are quite general but  are statistically
independent. On the other hand, we may think of the case where the channels are not necessarily statistically 
independent. This problem is quite hard in general. A typical tractable example of such networks would be a class of acyclic deterministic relay networks with no interference (called the Aref network)
in which the concept of ``channel capacity" is irrelevant. In this connection,  Ratnakar and
 Kramer  \cite{rat-kramer} have studied Aref networks with a single source and multiple  sinks, while 
Korada and Vasudevan \cite{korada-vasu} have studied Aref networks with multiple correlated sources and multiple sinks. The network capacity   formula as well as the network matching  formula obtained by them are in nice correspondence with the formula 
obtained by Ahlswede {\em et al.} \cite{al-yeung} as well as Theorem \ref{teiri:newji-1}
established in this paper, respectively.
}
\QED
\echui
\section*{Acknowledgments}
@ The author is very grateful to Prof. Shin'ichi Oishi for  
providing him with  pleasant research facilities during this work.   
%Also, thanks go to all  the colleagues with the Oishi laboratory for their kind helps.
%In addition,
 Thanks are   also due to Dinkar Vasudevan for bringing reference
 \cite{korada-vasu} to the author's attention.

%

%  \begin{figure}[htbp]
 %  \centering
%   \includegraphics[width=2in]{saddle.pdf} 
%   \includegraphics[bb= 0 0 200 220, scale=1.0]{fig1.pdf} 
 %  \includegraphics[bb= 0 0 200 420, scale=0.5]{multi-zumen0-counter} 
  % \caption{Counter example}
%\end{figure}

%
%

%
\end{document}